\documentclass[reprint,amsmath,amssymb,aps,pra,superscriptaddress]{revtex4-2}

\usepackage{upgreek}
\usepackage{xcolor}
\newcommand{\tr}{{\rm tr}}

\usepackage{graphicx}
\usepackage{dcolumn}
\usepackage{soul}
\usepackage{bm}

\begin{document}

\preprint{APS/123-QED}

\title{
Quantum Schr\"odinger bridges:\\ large deviations and time-symmetric ensembles
}

\author{Olga Movilla Miangolarra}
\affiliation{Departamento de Física, Universidad de La Laguna, La Laguna 38203, Spain}
\affiliation{Instituto Universitario de Estudios Avanzados (IUdEA), Universidad de La Laguna, La Laguna 38203, Spain}
\author{Ralph Sabbagh}
\affiliation{Department of Mechanical and Aerospace Engineering, University of California, Irvine, California 92697, USA}
\author{Tryphon T. Georgiou}
\affiliation{Department of Mechanical and Aerospace Engineering, University of California, Irvine, California 92697, USA}

\begin{abstract}
Quantum counterparts of Schr\"odinger's classical bridge problem have been around for the better part of half a century. During that time, several quantum approaches to this multifaceted classical problem have been introduced. In the present work, we unify, extend, and interpret several such approaches through a classical large deviations perspective. 
To this end, we consider time-symmetric ensembles that are pre- and post-selected before and after a Markovian experiment is performed. The Schr\"odinger bridge problem is that of finding the most likely joint distribution of initial and final outcomes that is consistent with obtained endpoint results. 
 The derived distribution provides quantum Markovian dynamics that bridge the observed endpoint states in the form of density matrices.
The solution retains its classical structure in that density matrices can be expressed as the product of forward-evolving and backward-evolving matrices. 
In addition, the quantum Schr\"odinger bridge allows inference of the most likely distribution of outcomes of an intervening measurement with unknown results. This distribution may be written as a product of forward- and backward-evolving expressions, in close analogy to the classical setting, and in a time-symmetric way.
The derived results are illustrated through a two-level amplitude damping~example.
\end{abstract}


\maketitle
\section{Introduction}

The early years of quantum mechanics were marked by intense debate and a search for a coherent interpretation of the new theory. This period of intellectual turmoil saw the founders of the theory grappling with the counterintuitive nature of quantum phenomena.
Amidst this turbulent backdrop, Erwin Schr\"odinger, already known for his wave mechanics formulation of quantum theory, sought an intuitive interpretation that led to significant, yet unexpected, contributions in the early 1930s \cite{schrodinger1931umkehrung,schrodinger1932theorie,chetrite2021schrodinger}.

Specifically, Schr\"odinger posed and solved the following problem in the setting of classical stochastic processes.
 Suppose we have an initial distribution of particles $p_0$ which are known to obey Brownian dynamics, i.e., their distribution follows a heat equation.  Assume an assistant observes the position of the finite, but large, number of particles at all times $t\in[0,1]$ without reporting their results.  At time $t=1$, we measure the empirical distribution of the particles, $p_1$, and find that it does not match the prediction obtained by solving the heat equation forward in time starting from the known initial distribution $p_0$. Clearly, something unlikely must have happened,  but what? 
 What is the most likely evolution of the particles observed by our assistant between times $0$ and $1$?

Schr\"odinger found that the structure of the solution to this problem had a striking similarity to quantum mechanics \cite{schrodinger1932theorie}. Indeed,  the most likely distribution of the  particles at intervening times is given by
$$
\varphi(t,x)\hat\varphi(t,x),
$$
where $\varphi$ and $\hat\varphi$ satisfy the forward and backward heat equations,
\begin{align}\label{eq:heat}
\partial_t\hat\varphi&=\frac12\Delta\hat\varphi\mbox{~~ and ~~}
    \partial_t\varphi=-\frac12\Delta\varphi,
\end{align}
respectively.
Analogously, the probability for a quantum mechanical particle to be found at position $x$ and time $t$ can be computed as
$$
\psi(t,x)\psi^*(t,x),
$$
where $\psi$ and $\psi^*$ solve Schr\"odinger's equation, which in its simplest form (no potential energy) reads
\begin{equation}\label{eq:schrod}
   \frac{1}{i\hbar} \partial_t\psi=\frac12\Delta \psi
\mbox{~~ and ~~}
\frac{1}{i\hbar} \partial_t\psi^*=-\frac12\Delta \psi^*, 
\end{equation}
where we have displayed the evolution of the complex conjugate for comparison. As a consequence of this analogy, Schr\"odinger concludes his series of lectures at the Institut Henri Poincaré by posing the following question~\cite{schrodinger1932theorie}: 
 should we describe quantum phenomena, not by the complex value of a wavefunction at one instant of time, but by a real probability at two different instances of time? 

Motivated by this analogy, Féynes~\cite{fenyes1952wahrscheinlichkeitstheoretische} and  Nelson~\cite{nelson2020dynamical}, among others,  sought to ground quantum mechanics on the classical theory of stochastic processes. 
They developed the framework of ``stochastic mechanics", where quantum particles follow stochastic paths. Which particular path is taken by the quantum particle is unknown to the observer, thus preserving the uncertainty principle.
In parallel, the classical probability problem posed by Schr\"odinger, that of finding the most likely dynamics that bridge two endpoint distributions, led to the birth of a new branch in probability -- large deviations theory. 

\pagebreak

In this context, Schr\"odinger's problem has been coined the \emph{Schr\"odinger bridge problem.}
It has recently captured a considerable amount of attention, not only for its diverse applications \cite{tong2020trajectorynet,somnath2023aligned,movilla2024inferring,miangolarra2024maximum}, but also for its different interpretations \cite{pavon2021data,leonard2012schrodinger}. Indeed, while the Schr\"odinger bridge problem can be seen as a large deviations problem, it can also be viewed as an inference problem in which one seeks to find a dynamical model that is maximally non-commital to unavailable information~\cite{pavon2021data}.
In addition, the Sch\"odinger bridge problem can be cast as both, 
the control problem to steer particles from one endpoint distribution to another through a  minimum energy control~\cite{chen2016relation}, and as an entropic regularization of the optimal transport problem~\cite{leonard2012schrodinger}. Part of the recent popularity of the Sch\"odinger bridge is due to its ease of computation through a globally convergent algorithm known as Fortet-Sinkhorn~\cite{chen2021stochastic}.
    
    In recent years, diverse quantum counterparts to the Schr\"odinger bridge problem have emerged. The endeavor was initiated in the fairly unknown work of Otto Bergmann~\cite{bergmann1988quantum}, where Schr\"odinger's solution is applied ``(unauthorized)" to quantum dynamics as an ``exercise".
 A decade later, the problem was taken on in several works~\cite{pavon2002quantum,beghi2002steer} that used Nelson's stochastic mechanics to rigorously apply the classical Schr\"odinger bridge solution to a quantum setting.  However, neither of these works~\cite{bergmann1988quantum,pavon2002quantum,beghi2002steer} was able to link the obtained solution to a most likely (or entropy minimizing) solution. It was not until 2010 that a step in that direction was made; the work \cite{pavon2010discrete} introduced new Schr\"odinger bridges as the solution to an entropy minimization problem over quantum state trajectories. 
  Nevertheless, the authors in \cite{pavon2010discrete} were only able to solve the half-bridge problem, i.e., that of matching an initial (or final) condition that is unexpected, as opposed to  reconciling both initial and final.  
  The problem to reconcile both, initial and final density matrices, across a quantum channel modeled by a Kraus map, was considered in~\cite{georgiou2015positive}, where a quantum counterpart to the Fortet-Sinkhorn algorithm was sought in the absence of a suitable optimization criterion.

 In the present work, we unify previous approaches to quantum Schr\"odinger bridges \cite{bergmann1988quantum}, \cite{pavon2010discrete} and \cite{georgiou2015positive}, 
 by developing a Schr\"odinger bridge (that matches both ends), as an extension to the half-bridges introduced in~\cite{pavon2010discrete}. In doing so, we provide a physical interpretation of the problem as a large deviations problem where the dynamics of the system are dictated by quantum mechanics. 
 The obtained solution bridges the two observed endpoint states
 in a way anticipated in
 the earlier work~\cite{georgiou2015positive}.
  Moreover, the endpoint states are shown to maintain the product structure of the classical solution,
which is carried over to the probability of a certain intermediate outcome, in the context of time-symmetric quantum measurement theory, as well as to its weak values.  Further, the time-symmetry of the classical bridges is shown to be preserved in this quantum setting.


\section{Pre- and post-selected quantum Schr\"odinger bridges}
\subsection{Experiment description}
\label{sec:exp}
\begin{figure*}[t!]
    \centering
    \includegraphics[width=6.5in]{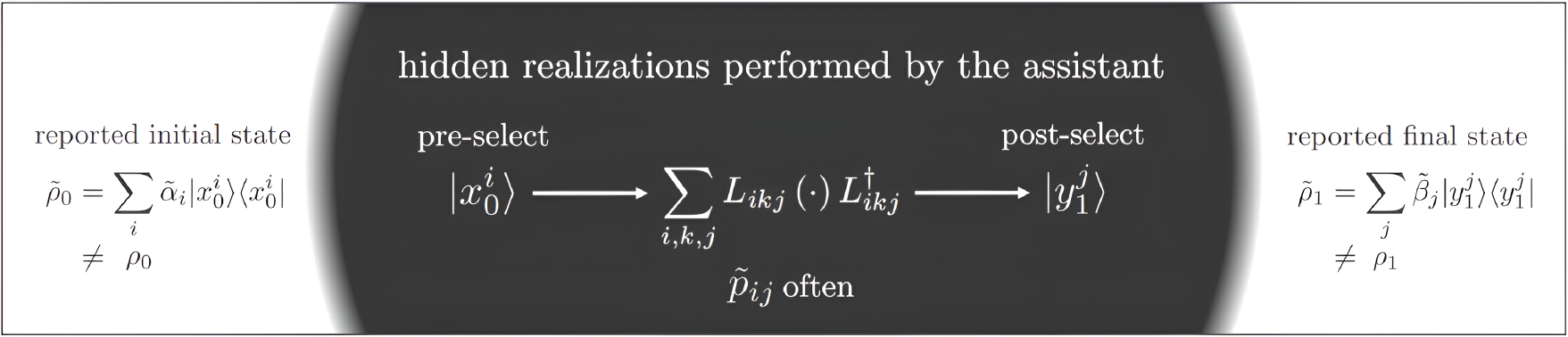}
    \caption{Quantum analog of Schr\"odinger's $1931$ bridge problem.}
    \label{fig:ds}
\end{figure*}
Let us introduce the following quantum analog of the problem that Schr\"odinger posed in 1931.
Consider a finite-dimensional quantum system.
Suppose we prepare an initial state, at time $t = 0$, by performing a projective measurement, with respect to a non-degenerate observable $X_0=\sum_ix_0^i|x_0^i\rangle\langle x_0^i|$, on a given state $\rho_{0^-}$. 
On average, we expect an initial mixed state characterized by the density matrix 
\begin{equation}\label{eq:priorrho0}
\rho_0=\sum_i\alpha_i|x_0^i\rangle\langle x_0^i|,
\end{equation}
where the probabilities $\alpha_i$ are prescribed by the state before the measurement.

The initial state 
$\rho_0$
is subjected to some known Markovian dynamics, during the interval of time $(0,1)$, described by the Kraus map 
$\sum_k L_k(1,0)(\cdot)L_k(1,0)^\dagger$. At time $t=1$, the experiment concludes with another projective measurement on the system of another (non-degenerate) observable $Y_1=\sum_j y_1^j|y_1^j\rangle\langle y_1^j|$. Therefore, the average (non-selective) dynamics of the system are given by 
\begin{align}
\sum_{i,k,j}L_{ikj}(1,0)(\cdot)L_{ikj}(1,0)^\dagger\label{eq:dynamics}
\end{align}
where, in a slight abuse of notation, the Kraus operator reads
\begin{equation}
    \label{eq:Likj}
    L_{ikj}(1,0)=\Pi^j_1L_k(1,0)\Pi^i_0,
\end{equation}
with $\Pi^i_0=|x_0^i\rangle\langle x_0^i|$ and $\Pi^j_1=|y_1^j\rangle\langle y_1^j|$ the projections onto particular initial and final states. 

Consequently, the average state of our system at the end of the experiment is expected to be
\begin{align}\nonumber
\rho_{1}&=\sum_{i,k,j}L_{ikj}(1,0)\rho_0 L_{ikj}(1,0)^\dagger
   \\
&=\sum_{i,j}p_{ij}|y_1^j\rangle\langle y_1^j|
=\sum_j\beta_j|y_1^j\rangle\langle y_1^j|, \label{eq:priorrho1}
\end{align}
where we have defined the joint probability of measuring $|x_0^i\rangle$ at time $t=0$ and $|y_1^j\rangle$ at time $t=1$ as
$$
p_{ij}=\sum_k\alpha_i|\langle y_1^j|L_k(1,0)|x_0^i\rangle|^2,
$$
and $\beta_j=\sum_i p_{ij}$ as the probability of obtaining a final measurement result associated to state $|y_1^j\rangle$. For simplicity, we will assume throughout that $\alpha_i,\beta_j,p_{ij}>0$.

Suppose we have an assistant who performs this experiment for us in the following way. Our assistant prepares an initial state through the described initial measurement on $\rho_{0^-}$ obtaining a certain state $|x_0^i\rangle$. Then, the initial state $|x_0^i\rangle$ is subjected to the prescribed Markovian dynamics~\eqref{eq:dynamics}. Finally, our assistant performs the final projective measurement with respect to states $\{|y_1^j\rangle\}$, leading to a certain final state $|y_1^j\rangle$.  In other words, our assistant realizes an experiment in which the states are \emph{{pre-  }and post-selected}.

After a large number of such experiments, say $N$, our assistant reports to us the fraction $\tilde\alpha_i$ of experiments that started with state $|x_0^i\rangle$ and the fraction $\tilde\beta_j$ of experiments that ended on state $|y_1^j\rangle$, without telling us which combination of states was chosen in each realization. That is, without reporting the joint distribution $\tilde p_{ij}$.  Thus, from our point of view, the experiment starts and concludes with the system in the mixed states
\begin{equation}\label{eq:endpoints}
\tilde\rho_0=\sum_i\tilde\alpha_i|x_0^i\rangle\langle x_0^i| \ \mbox{ and }\ 
\tilde\rho_1=\sum_j\tilde\beta_j|y_1^j\rangle\langle y_1^j|,
\end{equation}
 respectively, which do not align with our expectation $\rho_0$ and $\rho_1$. Moreover, the reported states are such that 
 $$
 \tilde\rho_1\neq \sum_{i,k,j}L_{ikj}(1,0)\tilde\rho_0L_{ikj}(1,0)^\dagger.
 $$
See Figure \ref{fig:ds} for a schematic representation of the experiment.

The discrepancy between the expected and the reported initial and final states may be due to large deviations, i.e., obtaining an unlikely ensemble from a collection of measurements due the finite size of the experimental record, or due to pre- and post-selection, where our assistant might have chosen a sub-ensemble of realizations to cook up some desired $\tilde\alpha_i$ and $\tilde\beta_j$ probabilities. 
Either way,  in the spirit of Schr\"odinger's original gedanken experiment, we pose the following question: what is the most likely joint probability $\tilde p_{ij}$ that led to the outcomes $\tilde\rho_0$ and $\tilde\rho_1$? In other words, if the outcomes were post-selected, what is the most likely way this post-selection was achieved?

\subsection{A large deviations solution}
The question raised above reduces to a \emph{classical} large deviations problem of the same nature as the one that Schr\"odinger answered in~\cite{schrodinger1931umkehrung}. In modern terms, we seek to quantify the likelihood of unexpected outcomes through Sanov's theorem, which states that the probability $P$ of drawing an atypical distribution from a finite collection of $N$ realizations decays exponentially as $N\to\infty$. Specifically, 
$$
P\sim e^{-IN },
$$
where the decay rate is given by the large deviations rate function $I$ that quantifies the distance between $P$ and the typical probability distribution. Thereby, the most likely atypical distribution that is consistent with given outcomes is the one that minimizes the rate function. 

Although the 
 experiment we consider involves a quantum evolution, the measurement at the two sites 
 where pre- and post-selection takes place,
render the probabilistic model of the experimental setting classical.
 As a result,
the rate function is the relative entropy
$$
\sum_{i,j}\tilde p_{ij}\log\frac{\tilde p_{ij}}{p_{ij}}, 
$$
 between the atypical observed distribution $\tilde p_{ij}$,  
and the expected (i.e., typical) one $p_{ij}$, which would have been obtained if no rare events or 
selection took place. 

Hence, we now seek the most likely (classical) joint probability distribution $\tilde p_{ij}$ between initial and final states $|x_0^{i}\rangle$ and $|y_1^{j}\rangle$ that is in agreement with the observation record, i.e.,
\begin{align}
\sum_j\tilde p_{ij}=\tilde \alpha_i~~\text{  and  }~~\sum_i\tilde p_{ij}=\tilde \beta_j.\label{eq:constraint}
\end{align}
This is nothing but 
a classical one-time-step Schr\"odinger bridge problem, to find the minimizer
\begin{align}\label{eq:minimization}
    {\rm arg}\min_{\tilde p_{ij}} \sum_{i,j}\tilde p_{ij}\log\frac{\tilde p_{ij}}{p_{ij}} ~~\text{   so that \eqref{eq:constraint}  holds.}
\end{align}
To solve this problem, we consider the augmented Lagrangian
$$
\mathcal L=  \sum_{i,j}\left(\tilde p_{ij}\log\frac{\tilde p_{ij}}{p_{ij}}+\lambda_i(\tilde p_{ij}-\tilde \alpha_i)+\gamma_j(\tilde p_{ij}-\tilde \beta_j)\right),
$$
where $\lambda_i$ and $\gamma_j$ are Lagrange multipliers.
Setting its first variation with respect to $\tilde p_{ij}$ to zero, we obtain that the optimal $\tilde p_{ij}$ can be expressed in the form
\begin{align}
\tilde p_{ij}^*=\frac{b_j}{a_i } \frac{\tilde\alpha_i}{\alpha_i}p_{ij},\label{eq:coupling}
\end{align}
where $a_i$ and $b_j$ must be chosen such that the constraints~\eqref{eq:constraint} are fulfilled. The minimizer $\tilde p_{ij}^*$ 
is the most likely joint probability, in the sense that it constitutes the empirical joint probability that produced the specified marginals with the highest chances of materializing in finite repetitions of the described experiment. In other words, such a joint probability dictates the {pre-  }and post-selection protocol, leading to the desired endpoint outcomes, that 
typically requires the least amount of  pre/post-selection, i.e., the least amount of interference by our assistant via {pre-  }and post-selection.

\subsection{The quantum bridge}
Putting these results back in our quantum dynamics, we find that
\begin{align*}\nonumber
   \tilde \rho_1 
&= \sum_{i,j}\tilde p_{ij}^*|y_1^j\rangle\langle y_1^j|
\\
&= \sum_{i,k,j}\frac{b_j}{a_i}\tilde \alpha_i|\langle y_1^j|L_k(1,0)|x_0^i\rangle|^2|y_1^j\rangle\langle y_1^j|
\\
&=\sum_{i,k,j}\frac{b_j}{a_i}\tilde \alpha_i\Pi_1^jL_k(1,0)\Pi_0^iL_k(1,0)^\dagger\Pi_1^j.
\end{align*}
Using the fact that $\{\Pi_0^i\}$ and $\{\Pi_1^j\}$ resolve the identity, i.e., ${\rm Id}=\sum_i \Pi_0^i=\sum_j \Pi_1^j$, and that they are orthogonal to each other in the sense that $\Pi_0^i\Pi_0^{i'}=\delta_{ii'}$ and $\Pi_1^j\Pi_1^{j'}=\delta_{jj'}$, with $\delta_{kl}$ the  Kronecker delta, we have that
\begin{align}
\nonumber
&\tilde \rho_1=\sum_{i,k,j}\Big(\sum_{j'}b_{j'}^{1/2}\Pi_1^{j'}\Big)L_{ikj}(1,0)\Big(\sum_{i''}a_{i''}^{-1/2}\Pi_0^{i''}\Big)\times\\
\nonumber
&\times\!\Big(\sum_{i'}\tilde \alpha_{i'}\Pi_0^{i'}\Big)\Big(\sum_{i'''}a_{i'''}^{-1/2}\Pi_0^{i'''}\Big)L_{ikj}(1,0)^\dagger\Big(\sum_{j''}b_{j''}^{1/2}\Pi_1^{j''}\Big)
\\
&\quad=\sum_{i,k,j} \tilde L_{ikj}(1,0)\tilde \rho_0\tilde L_{ikj}(1,0)^\dagger, \label{eq:newdyn}
\end{align}
where
\begin{equation}\label{eq:update}
\tilde L_{ikj}(1,0)=\upphi_1^{1/2}L_{ikj}(1,0)\upphi_0^{-1/2},
\end{equation}
with $\upphi_0$ and $\upphi_1$ given by 
\begin{subequations}
    \label{eq:Ssys}
\begin{align}
\upphi_0=\sum_ia_i|x_0^i\rangle\langle x_0^i|~~\mbox{ and }~~\upphi_1=\sum_jb_j|y_1^j\rangle\langle y_1^j|. \label{eq:phi's}
\end{align}

This notation is not arbitrary, indeed, we have that $\upphi_0$ is the evolution of $\upphi_1$ according to the \emph{adjoint} of the \emph{prior} Kraus map $L_{ikj}$, namely,
\begin{equation}
\label{eq:phidyn}
\upphi_0=\sum_{i,k,j}L_{ikj}(1,0)^\dagger\upphi_1L_{ikj}(1,0),
\end{equation}
since $\sum_j\tilde p^*_{ij}=\tilde \alpha_i$ implies $\sum_j b_jp_{ij}/\alpha_i ={a_i}$.  {We can interpret $\upphi$ as an observable in the Heisenberg picture which evolves backward in time, i.e., retrodictively \cite{wiseman2002weak}}.
Moreover, 
let 
\begin{equation}\label{eq:middle}
\hat\upphi_0=\upphi_0^{-1/2}\tilde \rho_0\upphi_0^{-1/2} \mbox{ and } \hat\upphi_1=\upphi_1^{-1/2}\tilde \rho_1\upphi_1^{-1/2},
\end{equation}
then, from \eqref{eq:newdyn} and \eqref{eq:update}, $\hat\upphi_1$ is the forward evolution  of $\hat\upphi_0$ under the prior Kraus map, i.e.,
\begin{equation}\label{eq:second}
\hat\upphi_1=\sum_{i,k,j}L_{ikj}(1,0)\hat\upphi_0L_{ikj}(1,0)^\dagger.
\end{equation}
\end{subequations}
Note that $\upphi_0,\,\upphi_1,\, \hat\upphi_0$, and $\hat\upphi_1$ are not density matrices; they are self-adjoint and positive definite
, but may not have trace $1$. Positivity follows under the mild assumption that $\alpha_i,\beta_j,p_{ij},\tilde\alpha_i,\tilde\beta_j>0$, in which case, $a_i>0$ and $b_j> 0$ as well.

Thus, we have identified in \eqref{eq:newdyn} a new Kraus map
$$
\sum_{i,k,j}\tilde L_{ikj}(1,0)(\cdot)\tilde L_{ikj}(1,0)^\dagger,
$$
that describes the most likely observed dynamics accounting for the discrepancy 
 between the expected $\rho_0, \rho_1$ and the observed $\tilde\rho_0, \tilde\rho_1$ 
 due to the finite nature of the experiment and consequent random effects (large deviations), and/or the pre/post-selection performed by our assistant. Indeed, it readily follows from \eqref{eq:phidyn} that $\sum_{i,j,k}\tilde L_{ikj}(1,0)^\dagger \tilde L_{ikj}(1,0)={\rm Id}$. Moreover, from \eqref{eq:middle}
 we have that $\tilde \rho$ at both ends of the experimental setting is the product of $\hat\upphi$ and $\upphi$, i.e.,
\begin{equation}
    \label{eq:for-rho}
    \tilde \rho_0=\upphi_0^{1/2}\hat\upphi_0\upphi_0^{1/2}\  \mbox{ and } \tilde \rho_1=\upphi_1^{1/2}\hat\upphi_1\upphi_1^{1/2}.
\end{equation}
where $\hat\upphi$ and $\upphi$ evolve according to the forward and adjoint prior dynamics, respectively, in analogy to the classical solution.

 Indeed, the system of equations \eqref{eq:Ssys}, represents a quantum counterpart of the classical Schr\"odinger system of coupled equations \cite{leonard2012schrodinger,chen2021stochastic,georgiou2015positive}.
The solution we have obtained, namely, the update of the dynamics
\eqref{eq:update}, is precisely of the form introduced in~\cite{georgiou2015positive}, applied, in this case, to Kraus operators of rank one ({pre-  }and post-selected). However, in the present work the structure of the solution as a modified Kraus map emerges from a classical large deviations experiment applied to the quantum setting via {pre-  }and post-selection, whereas in~\cite{georgiou2015positive} it was postulated as a counterpart of the Schr\"odinger-Fortet-Sinkhorn scheme applied to matrices. In addition and in contrast to~\cite{georgiou2015positive}, the existence of a solution to the Schr\"odinger system  \eqref{eq:Ssys} is guaranteed under the mild positivity assumptions stated, and the  solution can be obtained via a Fortet-Sinkhorn-like algorithm, cycling through  \eqref{eq:Ssys}  until convergence. We finally note that the form of the dynamical update, when applied to a half-bridge, is essentially that of \cite{pavon2010discrete}.

\subsection{Time-reversal of the quantum bridge}\label{sec:reversal}

Schr\"odinger's initial motivation centered on the fact that the adjoint of the heat equation in \eqref{eq:heat} coincides with its time-reversed counterpart, though the situation is a bit more nuanced when there is an advection term. Likewise, while the time-reversal of a unitary evolution (such as the one in the Schr\"odinger equation \eqref{eq:schrod}) is straightforwardly given by its adjoint, the time-reversal of general Kraus maps requires a suitable adjustment. In this section, echoing a similar property of the classical Schr\"odinger bridge, we illustrate that for a given Kraus map and marginal density matrices, the solution of the quantum Schr\"odinger bridge problem does not depend on the direction of time.

 To this end, we first note that Markovian dynamics, whether classical or quantum, can be time-reversed. For instance, starting from
\begin{align*}
    \rho_1 = \sum_{i,k,j}L_{ikj}(1,0)\rho_0L_{ikj}(1,0)^{\dagger},
\end{align*}
 with $\sum_{i,k,j}L_{ikj}(1,0)^{\dagger}L_{ikj}(1,0)=I$, that links $\rho_0$ to $\rho_1$,
a time-reversed Kraus map that links $\rho_1$ to $\rho_0$ is given by
\begin{align*}
    \rho_0 = \sum_{i,k,j}M_{ikj}(0,1)\rho_1M_{ikj}(0,1)^{\dagger},
\end{align*}
with 
\begin{equation}\label{eq:Mikj}
    M_{ikj}(0,1) = \rho_{0}^{1/2}L_{ikj}(1,0)^{\dagger}\rho_1^{-1/2}.
\end{equation}
Observe that $\sum_{i,k,j}M_{ikj}(0,1)^{\dagger}M_{ikj}(0,1)=I$; the requirement for $M_{ikj}(0,1)$ to constitute Kraus operators.
Both forward and time-reversed dynamics are equivalent in the sense that one can be recovered from the other. Note that here $\rho_1$ is invertible by assumption ($\beta_j>0$).

We can now solve the proposed quantum Schr\"odinger bridge problem with respect to the time-reversed dynamics. Specifically, the prior joint probability may be written with respect to the time-reversed dynamics as
\begin{align*}
    p_{ij}& = \beta_j\sum_{k}|\langle x_0^i|M_{ikj}(0,1)|y_1^j\rangle|^2 \\
    \bigg(&=
\alpha_i\sum_{k}|\langle y_1^j|L_{ikj}(1,0)|x_0^i\rangle|^2\bigg).
\nonumber
\end{align*}
We seek again the new joint probability $\tilde p_{ij}^*$ that minimizes relative entropy with respect to the prior
and satisfies the required marginals, i.e., that solves \eqref{eq:minimization}.
The optimal joint probability may be written as
\begin{align}\label{eq:sym-coupling}
    \tilde{p}^*_{ij}  = \cfrac{c_i}{d_j}\cfrac{\tilde{\beta_j}}{\beta_j}p_{ij}\ \bigg(= \cfrac{b_j}{a_i}\cfrac{\tilde{\alpha}_i}{\alpha_i}p_{ij}\bigg),
    \end{align}
where $c_i, d_j>0$ must be chosen such that constraints~\eqref{eq:constraint} are satisfied. Clearly, we have the following relations between the forward and reverse coefficients \begin{align}\tilde{\alpha}_i = a_ic_i\alpha_i,\ \mbox{ and }\ 
        \tilde{\beta}_j= b_jd_j\beta_j.\label{eq:abcd}
\end{align}

Following the same steps as before, and defining
\begin{align*}
\uppsi_0&=\sum_{i}c_i|x_{0}^i\rangle\langle x_{0}^i|,~~~~~~\uppsi_1=\sum_jd_j|y_{1}^j\rangle\langle y_{1}^j|,
\end{align*}
the updated time-reversed Kraus map takes the form
\begin{align}\label{eq:update-rev}
    \tilde{M}_{ikj}(0,1) = \uppsi_0^{1/2}M_{ikj}(0,1)\uppsi_1^{-1/2},
\end{align}
and bridges the observed marginals \eqref{eq:endpoints}, i.e., 
\begin{align}
    \tilde{\rho}_0 = \sum_{i,k,j}\tilde{M}_{ikj}(0,1)\tilde{\rho}_1\tilde{M}_{ikj}(0,1)^{\dagger}.\label{eq:rev}
\end{align}
{Moreover,} $\uppsi_1$ is related to $\uppsi_0$ through the adjoint map
\begin{align}
\uppsi_1=\sum_{i,k,j}M_{ikj}(0,1)^\dagger\uppsi_0 M_{ikj}(0,1).\label{eq:sys2}
\end{align}
Similarly, letting
\begin{align}
    \hat{\uppsi}_1 = \uppsi_1^{-1/2}\tilde{\rho}_1\uppsi_1^{-1/2}\text{   and   } \hat{\uppsi}_0 = \uppsi_0^{-1/2}\tilde{\rho}_0\uppsi_0^{-1/2},\label{eq:sys4}
\end{align}
then, from \eqref{eq:rev} and \eqref{eq:update-rev}, $\hat{\uppsi}_0$ is the evolution of $\hat{\uppsi}_1$ under the prior time-reversed map, namely,  
\begin{align}
    \hat{\uppsi}_0 = \sum_{i,k,j}M_{ikj}(0,1)\hat{\uppsi}_1M_{ikj}(0,1)^{\dagger}.\label{eq:sys3}
\end{align}
As before, $\uppsi_0,\,\uppsi_1,\, \hat\uppsi_0$, and $\hat\uppsi_1$ are not density matrices; they are self-adjoint and positive definite but need not have trace $1$. It readily follows from \eqref{eq:sys2} that $\sum_{i,j,k}\tilde M_{ikj}(0,1)^\dagger \tilde M_{ikj}(0,1)={\rm Id}$, and from \eqref{eq:sys4}
that $\tilde \rho$ at both ends of the experimental setting is the product of $\hat\uppsi$ and $\uppsi$, i.e.,
\begin{equation}\label{eq:rev-rho}
   \tilde \rho_0=\uppsi_0^{1/2}\hat\uppsi_0\uppsi_0^{1/2}\  \mbox{ and } \tilde \rho_1=\uppsi_1^{1/2}\hat\uppsi_1\uppsi_1^{1/2}. 
\end{equation}

Finally, we point out that the time-reversed bridge~\eqref{eq:rev} is equivalent to the forward bridge \eqref{eq:newdyn} up to time-reversal. That is, 
\begin{align}\label{eq:equiv}
    \tilde{M}_{ikj}(0,1) =   \tilde{\rho}_0^{1/2}\tilde{L}_{ikj}(1,0)^{\dag}\tilde{\rho}_1^{-1/2},
\end{align}
since 
$\uppsi_0^{1/2}M_{ikj}\uppsi_1^{-1/2}=   \tilde{\rho}_0^{1/2}\upphi_0^{-1/2}L^{\dag}_{ikj}\upphi_1^{1/2}\tilde{\rho}_1^{-1/2}.
$
This equality is true since, by \eqref{eq:abcd},
we have
\begin{equation}\label{eq:sym-rho}
\tilde{\rho}_0 =  \uppsi_0\rho_0 \upphi_0 \ \mbox{ and  } \ \tilde{\rho}_1 = \uppsi_1\rho_1 \upphi_1,  
\end{equation}
in analogy to the classical setting \cite[Eq. 36]{chen2016relation}. In fact, expression \eqref{eq:sym-rho} is time-symmetric, in the sense that both time directions are treated equally. Likewise, \eqref{eq:for-rho} and \eqref{eq:rev-rho} together with \eqref{eq:sym-rho} imply
\begin{equation*}
    \tilde{\rho}_0 =  \hat\uppsi_0\rho_0^{-1} \hat\upphi_0 \ \mbox{ and  } \ \tilde{\rho}_1 = \hat\uppsi_1\rho_1^{-1}\hat\upphi_1,
\end{equation*}
another time-symmetric expression.
We highlight that this last expression is written in terms of $\hat\upphi$ and $\hat\uppsi$, which evolve according to the forward and time-reversed prior Kraus maps (as opposed to their adjoints), respectively. These can be understood as unnormalized density matrices, since their traces are kept constant (but not necessarily equal to $1$) throughout the experiment.


\section{Quantum bridges with intervening projective 
measurements}

As the title of \cite{schrodinger1931umkehrung}, ``About the reversal of the laws of nature,''   suggests,  Schr\"odinger's bridge problem aimed at exploring and restoring time-symmetry in the conceptual framework of the quantum theory. Interestingly, a very similar aim underlies the theory by Aharonov, Bergmann, and Lebowitz~\cite{aharonov1964time},
to develop a time-symmetric formulation of quantum measurement. In this section, we draw a connection between the two.

\subsection{Most likely distribution of intervening projective measurement results}
\label{sec:weak}

We begin by recalling the two-state vector formalism discussed in~\cite{aharonov1964time}. Therein, the authors suggest that quantum measurement is not inherently asymmetric in time, but that the asymmetry comes from considering pre-selected ensembles only -- ensembles characterized by a common initial state. To build a time-symmetric theory, one needs to consider ensembles that treat both time directions in a symmetric manner, via {pre-  }and post-selected ensembles \footnote{The same is true for classical diffusion}.
Within this setting, they show that the probability of an intervening sequence of measurement outcomes, conditioned on pre- and post-selected states $|x_0\rangle$ and $|y_1\rangle$, is given by a time-symmetric expression~\cite{aharonov1964time}. In particular, for one intervening projective measurement of an observable $Z=\sum_zz|z\rangle\langle z|$ (and trivial dynamics), one obtains, using Bayes' theorem,
\begin{align}\label{eq:Aharanov}
P(z|x_0,y_1)
&=\frac{P(z,y_1|x_0)}{P(y_1|x_0)}=\frac{|\langle y_1|z\rangle\langle z|x_0\rangle|^2}{\sum_{z'} \!|\langle y_1|z'\rangle\langle z'|x_0\rangle|^2}.
\end{align}

More generally, we may consider a system that evolves according to a Markovian evolution specified over the two time-intervals $(0,\tau)$ and $(\tau, 1)$, where $0<\tau<1$, determined by Kraus maps with coefficients $K_l(\tau,0)$ and $K_{l'}(1,\tau)$, respectively. If the observable $Z$ is (ideally) measured at time $\tau$, then the probability of observing a (non-degenerate) eigenvalue $z$ in the {pre-  }and post-selected ensemble is
\begin{align}\label{eq:cond-prob}
P_\tau(z|x_0,y_1)
&=\!\frac{\sum_{l,l'}|\langle y_1|K_{l'}(1,\tau)|z\rangle\langle z|K_{l}(\tau,0)|x_0\rangle|^2}{\sum_{l,l'\!\!,\, z'} \!|\langle y_1|K_{l'}(1,\tau)|z'\rangle\langle z'|K_{l}(\tau,0)|x_0\rangle|^2}.
\end{align}
We note that \eqref{eq:cond-prob} need not be symmetric in time \footnote{Unlike the expression $|\langle y_1|z\rangle|^2$ in \eqref{eq:Aharanov}, the factor $\sum_{l'}|\langle y_1|K_{l'}(1,\tau)|z\rangle|^2$ in \eqref{eq:cond-prob} cannot be expressed in general as a backward transition probability.}. We also highlight that the normalization factor in  \eqref{eq:cond-prob} depends on both the initial and final states, making the overall dependence nonlinear.

Consider now a particular instance of the experiment described in Section \ref{sec:exp}, where the assistant performs a projective measurement of the observable $Z$ at time $\tau\in(0,1)$, without reporting its outcomes. The Markovian dynamics describing the experiment, between initial and final measurements, are given by the Kraus operator with coefficients
\begin{equation}
    \label{eq:priorx}
L_k(1,0)=K_{l'}(1,\tau)\Pi_z K_{l}(\tau,0),
\end{equation}
where $k=\{l,z,l'\}$ and $\Pi_z=|z\rangle\langle z|$. Once again, our assistant only reports the initial and final density matrices as in \eqref{eq:endpoints}, which do not match the ones we expected (eqs. \ref{eq:priorrho0} and \ref{eq:priorrho1}) from the Markovian dynamics \eqref{eq:priorx}. 

Following Schr\"odinger's dictum, we seek the most likely distribution
of measurement outcomes
\begin{align*}
\tilde P_\tau(z)=\sum_{i,j}\tilde p_{ij}P_\tau(z| y^j_1, x^i_0),
\end{align*}
obtained by our assitant at time $\tau$,
where $\tilde p_{ij}$ is the probability of starting at state $|x_0^i\rangle$ and ending at state  $|y_1^j\rangle$, as before. 
This expression is a standard disintegration of measure, where $\tilde p_{ij}$ is the \emph{coupling} and $P_\tau(z| y^j_1, x^i_0)$ is the bridge \footnote{It is not true for general quantum dynamics that probabilities can be disintegrated in this way; this is enabled by pre- and post-selection.}.
In our proposed problem, as in the classical Schr\"odinger bridge problem, the bridges $P_\tau(z| y^j_1, x^i_0)$ are known; they are fixed through \eqref{eq:cond-prob} by the prior dynamics~\eqref{eq:priorx}. Therefore, the problem of finding the most likely distribution reduces to finding, as before, the most likely coupling $\tilde p_{ij}$ recorded by the assistant during the experiment \footnote{The proof of this statement follows from a standard classical argument. For more details see \cite[Eq.\ 2]{georgiou2015positive}.}. Thus, the solution to this problem is given by \eqref{eq:coupling}, i.e.,
$$
\tilde p_{ij}^*=\frac{b_j}{a_i} \frac{\tilde\alpha_i}{\alpha_i}p_{ij},
$$
where $p_{ij}=\alpha_i\sum_k|\langle y_1^j|L_k(1,0)|x_0^i\rangle|^2=\alpha_i P(y_1^j|x_0^i),$ with $L_k(1,0)$ as in \eqref{eq:priorx}, and $a_i$ and $b_j$ such that \eqref{eq:constraint} is satisfied.

With this most likely coupling,  we can now infer the probability of obtaining an outcome $z$ at time $\tau\in(0,1)$, 
\begin{align}\nonumber
\tilde P^*_\tau(z)&=\sum_{i,j}\tilde p_{ij}^*\frac{P(z,y_1^j|x_0^i)}{P(y_1^j|x_0^i)}
\\\nonumber
&=\sum_{i,j,l',l}\tilde\alpha_i\frac{b_j}{a_i} |\langle  y_1^j|K_{l'}(1,\tau)\Pi_zK_{l}(\tau,0)| x_0^i\rangle|^2
\\\nonumber &
=\sum_{j,l'}b_j|\langle  y_1^j|K_{l'}(1,\tau)|z\rangle|^2\sum_{i,l}\frac{\tilde\alpha_i}{a_i}|\langle z|K_{l}(\tau,0)| x_0^i\rangle|^2
\\&=\varphi(\tau,z)\hat\varphi(\tau,z),
\label{eq:berg-prod}
\end{align}
where we have defined
\begin{subequations}\label{eq:varphi}
\begin{align}
\hat\varphi(\tau,z)&=\sum_{i,l}\frac{\tilde\alpha_i}{a_i}|\langle z|K_{l}(\tau,0)| x_0^i\rangle|^2,\\
 \varphi(\tau,z)&=\sum_{j,l'}b_j|\langle  y_1^j|K_{l'}(1,\tau)|z\rangle|^2.
\end{align}
\end{subequations}
Thus, the most likely probability \eqref{eq:berg-prod} decomposes into the product of two distributions, $\hat\varphi$ and $\varphi$, which may not be themselves normalized, but are such that their product is. Throughout, we assume that our experiment is designed so that
$\varphi(\tau,z),\hat\varphi(\tau,z)>0.$

It is instructive to consider the special case where the bases $\{|x_0^i\rangle\}$, $\{|y_1^j\rangle\}$ coincide with that of $Z$. Then, 
\begin{subequations}\label{eq:berg}
   \begin{align}
    \lim_{\tau\rightarrow 0^+}\tilde P^*_\tau(z_i)=\tilde\alpha_i,\\
    \lim_{\tau\rightarrow 1^-}\tilde P^*_\tau(z_j)=\tilde\beta_j,
\end{align} 
\end{subequations}
and thus the probability $\tilde P^*_\tau(z)$ interpolates, as a function of time $\tau$ \footnote{Of course, $\tilde P^*_\tau(z)$ is not a function of $\tau$ in the standard sense, since $\tau$ is assumed fixed throughout the experiment.}, the initial and final distributions, $\{\tilde \alpha_i\}$ and $\{\tilde \beta_j\}$, 
in the $Z$-eigenbasis. Interestingly, similar expressions to \eqref{eq:berg-prod} and \eqref{eq:berg} were obtained in \cite{bergmann1988quantum} through unwarranted analogies (differing only by normalization factors). Indeed, these are the closest quantum analogs of the classical Schr\"odinger bridge solution.

\subsection{The quantum bridge with intervening projective measurements}

Seeking a quantum analog of the classical bridge structure,
 the authors in \cite{georgiou2015positive}   postulated a way of decomposing Kraus maps into steps, where at intermediate times $t\in(0,1)$ the density matrix would be given by
\begin{equation}\label{eq:factoredform}
\tilde \rho_t=\upphi_t^{1/2}\hat\upphi_t\upphi_t^{1/2},
\end{equation}
with $\hat\upphi_t$ and $\upphi_t$ satisfying
$$
\hat\upphi_t=\sum_{i,k,j}L_{ikj}(t,0)\hat\upphi_0L_{ikj}(t,0)^\dagger
$$
and
$$
\upphi_t=\sum_{i,k,j}L_{ikj}(1,t)^\dagger\upphi_1L_{ikj}(1,t),
$$
respectively \footnote{{The Kraus map with coefficients $L_{ikj}(1,0)$ is assumed to be the composition of Kraus maps with coefficients $L_{ikj}(t,0)$, and $L_{ikj}(1,t)$, respectively.}}.  
However, in \cite{georgiou2015positive}
 no physical grounding was presented
 for the specific factored form~\eqref{eq:factoredform}. From the theory developed herein it follows that such a construction can be justified if one performs a projective measurement at those intermediate times, as we show in the following.

As before,  $\tilde\rho_1$ can be expressed in terms of a new Kraus map with coefficients~\eqref{eq:update}, where $\upphi_0$ and $\upphi_1$ are defined in \eqref{eq:phi's} and $L_{ikj}(1,0)$ in \eqref{eq:Likj} with $L_k(1,0)$ as in \eqref{eq:priorx}. In addition, $\tilde\rho$ at times $0$ and $1$ can be written as the product of $\upphi$ and $\hat\upphi$, which are again defined as before. This time, since our assistant performed a projective measurement at time $\tau$, we also have that
\begin{align*}
\nonumber\tilde\rho_\tau&=\sum_z\tilde P^*_\tau(z)|z\rangle \langle z| \\&=\sum_{i,j,l',l,z}\tilde \alpha_i\frac{b_j}{a_i} |\langle  y_1^j|K_{l'}(1,\tau)\Pi_zK_{l}(\tau,0)| x_0^i\rangle|^2|z\rangle\langle z|.
   \end{align*}
Following a similar computation as for \eqref{eq:newdyn}, we obtain
\begin{align}
      \tilde \rho_\tau&= \sum_{i,l,z}\tilde K_{ilz}(\tau,0)\tilde \rho_0\tilde K_{ilz}(\tau,0)^\dagger\label{eq:rhot}
\end{align}
where 
\begin{align*}
    \tilde K_{ilz}(\tau,0)&=\upphi_\tau^{1/2}K_{ilz}(\tau,0)\upphi_0^{-1/2}\\
    K_{ilz}(\tau,0)&=\Pi_zK_l(\tau,0)\Pi_0^i,
    \end{align*}
    and
    \begin{align*}
    \upphi_\tau&=\sum_z\varphi(\tau,z)|z\rangle\langle z|.
\end{align*}
Indeed, $\upphi_\tau$ is the backward evolution of $\upphi_1$ with respect to the adjoint of the prior dynamics up to time $\tau$, i.e.,
$$
\upphi_\tau=\sum_{z,l',j}K_{zl'j}(1,\tau)^\dagger\upphi_1 K_{zl'j}(1,\tau),
$$
with $$
K_{zl'j}(1,\tau)=\Pi_1^jK_{l'}(1,\tau)\Pi_z.
$$
Defining $\hat\upphi_\tau:=\upphi_\tau^{-1/2}\tilde\rho_\tau\upphi_\tau^{-1/2}$, equation \eqref{eq:berg-prod} implies
$$
\hat\upphi_\tau=\sum_z\hat\varphi(\tau,z)
|z\rangle\langle z|,
$$
while
\eqref{eq:rhot} leads to
$$
\hat\upphi_\tau=\sum_{i,l,z}K_{ilz}(\tau,0)\hat\upphi_0K_{ilz}(\tau,0)^\dagger,
$$
that is, $\hat\upphi_\tau$ is the forward evolution of $\hat\upphi_0$ under the prior dynamics.

We can now interpret the functions $\varphi(\tau,z)$ and $\hat\varphi(\tau,z)$ in terms of these matrices. Specifically, $\hat\varphi(\tau,z)/{\tr\{\hat\upphi_0\}}$ represents the probability of measuring $z$ at time $\tau$ given that we start at $\hat\upphi_0/\tr\{\hat\upphi_0\}$ --a normalized version of $\hat\upphi_0$. Thus, $\hat\varphi(\tau,z)$ represents the unnormalized probability of measuring $z$ in the forward evolved unnormalized density matrix $\hat\upphi_\tau$. The construction of $\varphi(\tau,z)$ is similar in the sense that $\varphi(\tau,z)=\tr\{\Pi_z\upphi_\tau\}$. However, it cannot be interpreted as an ``unnormalized probability", since $\upphi_1$ is evolved backward according to the adjoint Kraus map (which is not necessarily a Kraus map itself), and should therefore be regarded as an observable.

Bringing all together, we have that
$$
\tilde \rho_\tau=\upphi_\tau^{1/2}\hat\upphi_\tau\upphi_\tau^{1/2}
$$
is the product between the forward evolved $\hat\upphi_\tau$ and the backward evolved $\upphi_\tau.$
Moreover, the total updated Kraus map \eqref{eq:update} is the composition of the map taking $\tilde\rho_0$ to $\tilde \rho_\tau$, determined by the Kraus operator
\begin{subequations}
\label{eq:updatefor}
\begin{equation}
  \label{eq:update-for1}\upphi_\tau^{1/2}K_{ilz}(\tau,0)\upphi_0^{-1/2}, 
\end{equation}
and the map that takes $\tilde \rho_\tau$ into $\tilde \rho_1$ determined by
\begin{equation}
    \label{eq:update-for2}
\upphi_1^{1/2}K_{zl'j}(1,\tau)\upphi_\tau^{-1/2}.
\end{equation}
\end{subequations}
These results provide an extension to those presented in~\cite{pavon2010discrete} for the special case of a half-bridge. They represent a particular instance of the solution in~\cite{georgiou2015positive} for the case of an experiment with three projective measurements. The present work provides a physical experiment, as well as an interpretation in terms of the solution to an optimization problem. The results presented in this section can be readily extended to multiple intervening projective measurements.

\subsection{Time-reversal of the quantum bridge with intervening projective measurements}

As in Section \ref{sec:reversal}, we now consider the bridge problem with an intervening projective measurement in a time-reversed setting.
First, we define the 
prior intermediate state
$$
\rho_\tau=\sum_{z}P_\tau(z)|z\rangle\langle z|,
$$
where the prior probability of obtaining outcome $z$ at time $\tau$ is
$$
P_\tau(z)=\sum_{i,l}\alpha_i|\langle z|K_l(\tau,0)|x_0^i\rangle|^2,
$$
which is assumed to be strictly positive.
We time-reverse the Markovian evolution \eqref{eq:priorx} by separately reversing the evolution over the two intervals $(0,\tau)$ and $(\tau, 1)$. 
The resulting Kraus operators are
\begin{align*}
     N_{l'}(\tau,1)&=\rho_\tau^{1/2}K_{l'}(1,\tau)^\dagger\rho_1^{-1/2} \mbox{ and}\\
N_{l}(0,\tau)&=\rho_0^{1/2}K_{l}(\tau,0)^\dagger\rho_\tau^{-1/2},
\end{align*}
which satisfy, respectively,
\begin{align*}
\rho_\tau&=\sum_{l'}N_{l'}(\tau,1)\rho_1 N_{l'}(\tau,1)^\dagger\mbox{ and}
 \\\rho_0&=\sum_{l}N_{l}(0,\tau)
\rho_\tau
N_{l}(0,\tau)^\dagger.\nonumber
\end{align*}

Then, using \eqref{eq:sym-coupling} together with $p_{ij}=\beta_j P(x_0^i|y_1^j)$, and following the same steps as in the forward case, we have
\begin{align}\nonumber
\tilde P^*_\tau(z)&=\sum_{i,j}\tilde p_{ij}^*\frac{P(z,x_0^i|y_1^j)}{P(x_0^i|y_1^j)}\\\nonumber
&=\sum_{i,j,l',l}\tilde\beta_j\frac{c_i}{d_j} |\langle  x_0^i|N_{l}(0,\tau)\Pi_zN_{l'}(\tau,1)| y_1^j\rangle|^2
\\&=\psi(\tau,z)\hat\psi(\tau,z), \label{eq:rev-berg-prod}
\end{align}
where we have defined
\begin{align*}
\hat\psi(\tau,z)&=\sum_{j,l'}\frac{\tilde\beta_j}{d_j}|\langle z|N_{l'}(\tau,1)| y_1^j\rangle|^2,\\
\psi(\tau,z)&=\sum_{i,l}c_i|\langle  x_0^i|N_{l}(0,\tau)|z\rangle|^2.
\end{align*}

By direct substitution of the definition of $N_{l'}$ in \eqref{eq:rev-berg-prod}, together with \eqref{eq:abcd} and \eqref{eq:varphi}, we obtain that 
\begin{equation}
    \label{eq:Pz-rev1}
    \tilde P^*_\tau(z)=\psi(\tau,z)P_\tau(z)\varphi(\tau,z).
\end{equation}
Similarly, substituting for
$N_{l}$ instead of $N_{l'}$ we have
\begin{equation}\label{eq:Pz-rev2}
\tilde P^*_\tau(z)=\hat\varphi(\tau,z)P_\tau(z)^{-1}\hat\psi(\tau,z).
\end{equation}
which is also a time-symmetric expression. The advantage of this last expression is that the three elements can be interpreted as probabilities; $\hat\varphi$ and $\hat\psi$ being unnormalized and flowing in the forward and reverse directions, respectively.
In addition, by substituting both $N_l$ and $N_{l'}$  in \eqref{eq:rev-berg-prod}, together with \eqref{eq:abcd}, it is easy to check that \eqref{eq:rev-berg-prod} and \eqref{eq:berg-prod} are equal.

Thus, the updated time-reversed Kraus map is obtained by writing the intermediate state as 
\begin{align*}
\nonumber\tilde\rho_\tau&=\sum_z\psi(\tau,z)\hat\psi(\tau,z)|z\rangle \langle z| \\&= \sum_{j,l',z}\tilde N_{zl'j}(0,\tau)\tilde \rho_1\tilde N_{zl'j}(0,\tau)^\dagger
\end{align*}
where 
\begin{subequations}
    \label{eq:updaterev}
\begin{align}
    \label{eq:update-rev1}
\tilde N_{zl'j}(\tau,1)&=\uppsi_\tau^{1/2}N_{zl'j}(\tau,1)\uppsi_1^{-1/2},\\\nonumber
N_{zl'j}(\tau,1)&=\Pi_zN_{l'}(\tau,1)\Pi_1^j,
\end{align}
and 
$$
\uppsi_\tau=\sum_{z}\psi(\tau,z)|z\rangle\langle z|
$$
is the evolution of $\uppsi_0$ (defined in Section \ref{sec:reversal}) according to the adjoint of the prior time-reversed Kraus map.

Defining $\hat\uppsi_\tau:=\uppsi_\tau^{-1/2}\tilde\rho_\tau\uppsi_\tau^{-1/2}$,  we have that
$$
\tilde\rho_\tau=\uppsi_\tau^{1/2}\hat\uppsi_\tau\uppsi_\tau^{1/2}=\uppsi_\tau\rho_\tau\upphi_\tau=\hat\uppsi_\tau\rho_\tau^{-1}\hat\upphi_\tau,
$$
which follow from (\ref{eq:rev-berg-prod}-\ref{eq:Pz-rev2}).
Finally, since the total time-reversed update is given by \eqref{eq:update-rev}, we have that the Kraus map determined by the operators
\begin{equation}\label{eq:update-rev2}
    \tilde N_{ilz}(0,\tau)=\uppsi_0^{1/2}N_{ilz}(0,\tau)\uppsi_\tau^{-1/2},
\end{equation}
\end{subequations}
where
$$
N_{ilz}(0,\tau)=\Pi_0^iN_{l}(0,\tau)\Pi_z ,
$$
maps $\tilde\rho_\tau$ to $\tilde\rho_0$. It is easy to check, as we did in equation~\eqref{eq:equiv}, that these updated Kraus maps \eqref{eq:updaterev} are equivalent to \eqref{eq:updatefor} up to time reversal.


\section{Quantum bridges with intervening generalized 
measurements}

\subsection{Most likely distribution of intervening
generalized measurement results}

To infer distributions of outcomes of intervening measurements, it is not essential to have the intervening measurement be projective. Indeed, some of the results obtained in the last section can be extended 
to generalized intervening measurements described by a Kraus map with operators $\Omega_{\bar z}(\tau_2,\tau_1)$ (instead of $\Pi_z$ which would be associated with an instantaneous projective measurement). These operators capture generalized measurements of the observable $Z=\sum_zz\Pi_z$, with outcome $\bar z$,  that take place during the time interval $ (\tau_1,\tau_2)$, with $0<\tau_1<\tau_2<1$. 

Specifically, let us consider  the experiment described in Section \ref{sec:exp}, this time with Kraus operators
$$
L_k(1,0)=K_{l'}(1,\tau_2)\Omega_{\bar z}(\tau_2,\tau_1) K_{l}(\tau_1,0),
$$
with $k=\{l,\bar z,l'\}.$ After our assistant reports to us initial and final distributions that do not match our expectations, we seek to infer the most likely probability of obtaining an intervening result $\bar z$. For the described pre- and post-selected ensemble we have that this probability is given by
\begin{align*}
\tilde P_\Omega(\bar z)=\sum_{i,j}\tilde p_{ij}P_\Omega(\bar z| y^j_1, x^i_0)=\sum_{i,j}\tilde p_{ij}\frac{P_\Omega(\bar z,y^j_1| x^i_0)}{P_\Omega( y^j_1| x^i_0)}.
\end{align*}
Following the same argument as for the projective case, the most likely outcome distribution is determined by the  optimal coupling
$\tilde p^*_{ij}=\tilde\alpha_ib_j P_\Omega( y^j_1| x^i_0)/a_i$,
as
\begin{align}\nonumber
\tilde P_\Omega(\bar z)&=\sum_{i,j}b_j\frac{\tilde \alpha_i}{a_i}P_\Omega(\bar z,y^j_1| x^i_0)
\\\nonumber &=\sum_{i,j,l,l'}b_j\frac{\tilde \alpha_i}{a_i}|\langle y_1^j|K_{l'}(1,\tau_2)\Omega_{\bar z}(\tau_2,\tau_1) K_{l}(\tau_1,0) |x_0^i\rangle|^2
\\&=\tr \big\{\upphi_{\tau_2} \Omega_{\bar z}(\tau_2,\tau_1)\hat\upphi_{\tau_1}\Omega_{\bar z}(\tau_2,\tau_1)^\dagger\big\},\label{eq:generalized-distrib}
\end{align}
and where $\upphi_{\tau_2}$ and $\hat\upphi_{\tau_1}$ are defined, similarly to before, as
\begin{align*}
\upphi_{\tau_2}&=\sum_{l'}K_{l'}(1,\tau_2)^\dagger\upphi_{1}K_{l'}(1,\tau_2), \mbox{ and }
\\
\hat\upphi_{\tau_1}&=\sum_{l}K_{l}(\tau_1,0)\hat\upphi_{0}K_{l}(\tau_1,0)^\dagger.
\end{align*}
 In contrast to \eqref{eq:berg-prod}, the expression \eqref{eq:generalized-distrib} for the distribution of generalized measurements cannot be expressed, in general, as a product of forward and backward evolving functions, in complete analogy to the classical setting. 

{This distribution of outcomes can be similarly obtained in terms of the time-reversed maps
\begin{subequations}\label{eq:reverse-maps}
  \begin{align}
     N_{l'}(\tau_2,1)&:=\rho_{\tau_2}^{1/2}K_{l'}(1,\tau_2)^\dagger\rho_{1}^{-1/2},
     \\
     \Omega_{\bar z}^{\rm rev}(\tau_1,\tau_2)&:=\rho_{\tau_1}^{1/2}\Omega_{\bar z}(\tau_2,\tau_1)^\dagger\rho_{\tau_2}^{-1/2},
     \\
N_{l}(0,\tau_1)&:=\rho_0^{1/2}K_{l}(\tau_1,0)^\dagger\rho_{\tau_1}^{-1/2},
\end{align}  
\end{subequations}
where we have defined the prior states
\begin{align*}
    \rho_{\tau_1}&=\sum_{l} K_l(\tau_1,0)\rho_0 K_l(\tau_1,0)^\dagger\\
    \rho_{\tau_2}&=\int_{\mathbb R}\Omega_{\bar z}(\tau_2,\tau_1)\rho_{\tau_1}\Omega_{\bar z}(\tau_2,\tau_1)^\dagger d\bar z.
\end{align*}
Specifically, we have that, similarly to the projective case,
\begin{align*}
\tilde P_\Omega(\bar z)&=\sum_{i,j}c_i\frac{\tilde \beta_j}{d_j}P_\Omega(\bar z,x^i_0| y^j_1)
\\&=\!\!\sum_{i,j,l,l'}\!\!c_i\frac{\tilde \beta_j}{d_j}|\langle x_0^i|N_{l}(0,\tau_1)\Omega^{\rm rev}_{\bar z}(\tau_1,\tau_2) N_{l'}(\tau_2,1) |y_1^j\rangle|^2
\\&=\tr \big\{\uppsi_{\tau_1} \Omega^{\rm rev}_{\bar z}(\tau_1,\tau_2) \hat\uppsi_{\tau_2}\Omega^{\rm rev}_{\bar z}(\tau_1,\tau_2) ^\dagger\big\},
\end{align*}
where $\uppsi_{\tau_1}$ and $\hat\uppsi_{\tau_2}$ are given by 
 \begin{align*}
\uppsi_{\tau_1}&=\sum_{l}N_{l}(0,\tau_1)^\dagger\uppsi_{0}N_{l}(0,\tau_1), \mbox{ and }
\\
\hat\uppsi_{\tau_2}&=\sum_{l'}N_{l'}(\tau_2,1)\hat\uppsi_{1}N_{l'}(\tau_2,1)^\dagger.
 \end{align*}
Substituting in the definition of the time-reversed operators \eqref{eq:reverse-maps}, we clearly obtain the forward expression~\eqref{eq:generalized-distrib}.}

Thus, we have obtained the most likely distribution of observed outcomes of the measurements described by~$\Omega_{\bar z}$, with a structure that now slightly deviates from the classical structure.
Moreover, as these generalized measurements do not in general collapse the state onto a particular basis (as in the projective case), the basis at the intervening times of the measurement is not fixed. Therefore, the joint probability of initial, intervening, and final outcomes is no longer a classical Markovian joint probability, and
splitting the obtained bridge into any two (or three) steps is not justified. That is, the most likely Kraus map that takes $\tilde\rho_0\to\tilde\rho_1$ is still determined by  \eqref{eq:update}, but its splitting into steps \eqref{eq:update-for1} and \eqref{eq:update-for2} can no longer be justified.

\subsection{Most likely weak value}

A particularly interesting instance of a generalized measurement is that of a weak measurement~\cite{jacobs2006straightforward}.  Consider a quantum evolution determined by the Kraus operators
    \begin{equation}
        \label{eq:weak-kraus}
        L_k(1+\delta,0)= K_{l'}(1,\tau)\Omega^\delta_{\bar z}K_{l}(\tau,0),
    \end{equation}
where the generalized measurement operator $\Omega^{\delta}_{\bar z}$ is given by 
$$
\Omega^{\delta}_{\bar z}={\left(\frac{\delta}{2\pi}\right)^{1/4}}\sum_z e^{-\frac{\delta}{4}(z-\bar z)^2}\Pi_z,$$
with $\delta >0$ denoting the length of the interval of time during which the measurement takes place.  
In the limit as $\delta\to 0$, the variance of the Gaussian goes to infinity, and the information obtained from this generalized measurement tends to zero. 
Thus, in the small $\delta$ limit, this operator describes a weak measurement~\cite{jacobs2006straightforward}, where each single measurement is so uninformative that it does not affect the system's dynamics on average. 
In a slight abuse of notation, we have kept the indexing of the Kraus operators $K_{l'}$ as if the measurement took place instantaneously. This is done to emphasize the fact that both $K_l$ and $K_{l'}$ are independent of the duration of the intervening measurement.

These weak measurements are particularly interesting in the context of pre- and post-selected ensembles, since they turn out to give average values well beyond the admissible range of outcomes associated with observable $Z$~\cite{aharonov1988result,wiseman2002weak}. Specifically, weakly measuring $Z$ in an ensemble that is pre-selected at state $|x^i_0\rangle$ and post-selected at state $|y^j_1\rangle$, 
 leads to the average value (see the Appendix for more details),
 \begin{subequations}
\begin{equation}
Z^{ij}_W={\rm Re}\frac{\tr\big\{\sigma^j_\tau Z\hat\sigma^i_\tau\big\}}{\tr\{\sigma^j_\tau\hat\sigma^i_\tau\}},\label{eq:weakval}
\end{equation}
coined as \emph{weak value} \cite{aharonov1988result}.
Here, we have defined the backward-evolving observable 
\begin{equation}\label{eq:sig}
    \sigma^j_\tau:=\sum_{l'}K_{l'}(1,\tau)^\dagger|y^j_1\rangle\langle y^j_1|K_{l'}(1,\tau),
\end{equation}
and the forward-evolving density matrix 
\begin{equation}\label{eq:sighat} \hat\sigma^i_\tau:=\sum_lK_l(\tau,0)|x^i_0\rangle\langle x^i_0|K_l(\tau,0)^\dagger.\phantom{xx}
\end{equation}
 \end{subequations}
By carefully selecting the initial and final states $|x^i_0\rangle$ and $|y^j_1\rangle$, $Z_W$ can be made arbitrarily large \cite{aharonov1988result}.


Let us now consider the experiment described in Section~\ref{sec:exp}, associated to Kraus operators \eqref{eq:weak-kraus} with $\delta\to 0$, where our assistant reported obtaining initial and final average states $\tilde\rho_0$ and $\tilde\rho_1$ as in \eqref{eq:endpoints}. 
We seek to infer the most likely weak value
\begin{align*}
\tilde Z_W=\sum_{i,j}\tilde p_{ij} Z^{ij}_W
\end{align*}
 obtained by our assistant. As before, the most likely such value is given by the optimal coupling $\tilde p^*_{ij}=b_j\tilde \alpha_ip_{ij}/(\alpha_ia_i)=\tr\{\sigma_\tau^j\hat\sigma_\tau^i\}b_j\tilde\alpha_i/a_i$, i.e.,
\begin{align}\nonumber
\tilde Z^*_W&=\sum_{i,j}\tilde p^*_{ij} Z^{ij}_W\\&={\rm Re} \big\{\tr\big(\upphi_\tau Z\hat\upphi_\tau\big)\big\},\label{eq:weak-bridge}
\end{align}
where $\hat\upphi_\tau$ is the forward evolution of $\hat\upphi_0$, i.e.,
$$
\hat\upphi_\tau=\sum_{l}K_l(\tau,0)\hat\upphi_0K_l(\tau,0)^\dagger,
$$
and $\upphi_\tau$ is the backward evolution of $\upphi_1$ according to the adjoint Kraus map, this is,
$$
\upphi_\tau=\sum_{l'}K_{l'}(1,\tau)^\dagger\upphi_1 K_{l'}(1,\tau).
$$
 It is interesting to note that, 
thanks to the unnormalized matrices $\hat\upphi_\tau$ and $\upphi_\tau$, the most likely weak value loses its nonlinear normalization factor, as we have seen for the other types of measurements.

Finally, we explain how to build a weak bridge based on time-reversed dynamics. 
To this end, let us define, similarly to before, the time-reversed Kraus operators
\begin{align*}
     N_{l'}(\tau,1)&=\rho_\tau^{1/2}K_{l'}(1,\tau)^\dagger\rho_1^{-1/2} \mbox{ and}\\
N_{l}(0,\tau)&=\rho_0^{1/2}K_{l}(\tau,0)^\dagger\rho_\tau^{-1/2},
\end{align*}
 where, this time, $\rho_\tau$ is given by
$$
\rho_\tau=\sum_lK_l(\tau,0)\rho_0K_l(\tau,0)^\dagger,
$$
with no projection at time $\tau$. In addition, define
$$
Z^{\rm rev}=\rho_\tau^{1/2}Z\rho_\tau^{-1/2}
$$
to account for the time-reversal of the weak measurement operation.

Then, as is shown in the Appendix, the weak value obtained, by pre/post-selecting states $|x^i_0\rangle$ and $|y^j_1\rangle$,  can be expressed through these time-reversed maps as 
\begin{subequations}
\begin{align}\label{eq:weakval-rev}
Z^{ij,\rm rev}_W={\rm Re}\frac{\tr\big\{\varsigma_\tau^i Z^{\rm rev}\hat\varsigma_\tau^j\big\}}{\tr\{\varsigma_\tau^i\hat\varsigma_\tau^j\}},
\end{align}
where
\begin{equation}
\hat\varsigma_\tau^j=\sum_{l'}N_{l'}(\tau,1)|y_1^j\rangle\langle y_1^j|N_{l'}(\tau,1)^\dagger,
\end{equation} 
denotes the evolution of the post-selected state according to the prior time-reversed Kraus map, and
\begin{equation}
    \varsigma_\tau^i=\sum_{l}N_{l}(0,\tau)^\dagger|x_0^i\rangle\langle x_0^i|N_{l}(0,\tau)
\end{equation}
denotes the evolution of the pre-selected state according to the adjoint of the prior time-reversed map.
\end{subequations}



Once again, the most likely average $\sum_{i,j}\tilde p_{ij} Z^{ij,\rm rev}_W$ obtained by our assistant is determined by the optimal coupling $\tilde p_{ij}^*=c_i\tilde\beta_j \tr\{\varsigma_\tau^i\hat\varsigma^j_\tau\}/d_j$. Thus, the most likely weak value reads
\begin{align}\nonumber \label{eq:weak-rev}
\tilde Z^*_W&=\sum_{i,j}\tilde p^*_{ij} Z^{ij,\rm rev}_W\\&={\rm Re}\big\{\tr\big(\uppsi_\tau Z^{\rm rev}\hat\uppsi_\tau\big)\big\},
\end{align}
where $\hat\uppsi_\tau$ is the time-reversed evolution of $\hat\uppsi_1$, i.e.,
$$
\hat\uppsi_\tau=\sum_{l'}N_{l'}(\tau,1)\hat\uppsi_1N_{l'}(\tau,1)^\dagger,
$$
and $\uppsi_\tau$ is the evolution of $\uppsi_0$ according to the adjoint time-reversed Kraus map, this is,
$$
\uppsi_\tau=\sum_{l}N_l(0,\tau)^\dagger\uppsi_0N_l(0,\tau).
$$
It is straightforward to see that, substituting in \eqref{eq:weak-rev} the definition of $N_l(0,\tau)$, $Z^{\rm rev}$, and $N_{l'}(\tau,1)$, the weak values in the forward and time-reversed bridges coincide.

\section{Amplitude damping example}

Let us illustrate the obtained results through a simple example. In particular,
consider a two-state system such as a spin-1/2 particle or a two-level atom. Assume our assistant prepares the system at a spin $x$ up (down) state with probability $1/3$ (2/3), this is, 
$$
\tilde \rho_0=\frac23|0\rangle_x\langle 0|+\frac13|1\rangle_x\langle 1|.
$$
Then, our assistant performs an amplitude damping experiment in the $z$ direction, during which state $|1\rangle_z$ decays to $|0\rangle_z$ with some probability $\lambda$ due to dissipative effects. Specifically, in the $z$ eigenbasis we have
$$
K_0(t_2,t_1)=\left(\begin{array}{cc}
    1 &  0\\
    0 & \sqrt{1-\lambda(t_2-t_1)}
\end{array}\right)$$
and $$K_1(t_2,t_1)=\left(\begin{array}{cc}
    0 &  \sqrt{\lambda(t_2-t_1)}\\
    0 & 0
\end{array}\right),
$$
where $\lambda(t)=1-e^{-\gamma t}$ for some positive decay constant~$\gamma$.
Let us assume that at some instant of time $\tau$, our assistant performs an intervening projective measurement onto the $z$ eigenbasis, and that at the final time $t=1$, the state is found to be at
$$
\tilde \rho_1=\frac34|0\rangle_z\langle 0|+\frac14|1\rangle_z\langle 1| 
$$
on average,
through another $z$-eigenbasis measurement. Therefore, the total Kraus operators corresponding to these dynamics read
$$
L_{xkz}(1,0)=\Pi_{z}K_{l'}(1,\tau)\Pi_{z'}K_{l}(\tau,0)\Pi_{x},
$$
for $k=\{l,z',l'\}$ with $l,l',x,z,z'\in\{0,1\}$.
However, we find that
$$
\rho_1=\sum_{x,k,z}L_{xkz}(1,0) \rho_0L_{xkz}(1,0)^\dagger\neq\tilde \rho_1,
$$
where $\rho_0=\tilde\rho_0$;
that is, the observed endpoints do not match what we expected.

\begin{figure}[tb]
    \centering
\includegraphics[width=0.85\linewidth,trim={0.6cm 1.45cm 1.4cm 1.5cm},clip]{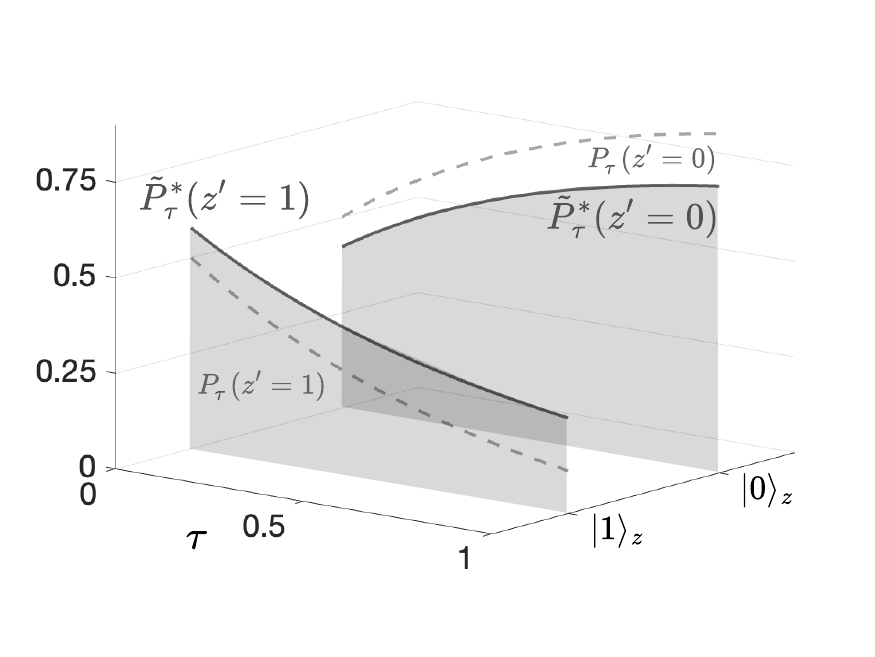}
    \caption{Probability $P_\tau(z')$ of measuring $z'\in\{0,1\}$ at different times $\tau$. 
    The inferred most likely probability is plotted in solid, while the dashed line captures the prior probability.
    The parameter $\gamma$ has been set to $1.5$.}
    \label{fig:qubit}
\end{figure}

Then, the most likely joint probability of observing particular $x$ and $z$ outcomes is given by \eqref{eq:coupling}, where $a_i$ and $b_j$ (two constants each) can be simply computed through the four algebraic equations \eqref{eq:constraint} that constitute our constraints~\footnote{For higher dimensional cases a simple Fortet-Sinkhorn algorithm can be implemented to compute these values.}.
The effective Kraus map that led to the observed endpoints is given by \eqref{eq:update}, while the intervening $z$-measurement results can be inferred through \eqref{eq:berg-prod}. The probability $\tilde P_\tau^{*}(z')$ is plotted solid in Figure \ref{fig:qubit} for all possible times $\tau\in(0,1)$, together with the prior probability $P_\tau(z')$ which is depicted by the dashed line for comparison. It is seen that post-selection affects $\tilde P_\tau^*(z')$ even as $\tau\to 0$; this would not have been the case if at time $t=0$ a $z$-eigenbasis measurement (instead of $x$) had been performed.
 Note that $P_\tau(z')$ should not be understood as a standard function of time since $\tau$ is fixed and unique throughout the experiment; i.e., only one intervening measurement is actually performed. 
 


\section{Conclusions}
In these pages we have proposed, developed, and illustrated a quantum Schr\"odinger bridge based on large deviations and pre- and post-selection. In doing so, we have separated the classical aspect of the problem from the underlying quantum features. Specifically, we have seen how measurements at two instances of time allow us to define classical joint probabilities through which a large deviations principle can be built. The solution to this classical problem leads to the effective quantum dynamics of the underlying system, as well as the inferred intermediate quantum measurement results, whether weak or strong. The structure of the solutions is analogous to their classical counterpart in that the most likely density matrices, as well as the intervening measurement distributions, turn out to be a product of forward and backward-evolving expressions. Furthermore, these expressions may be written in a time-symmetric way, as in the classical setting.

The proposed problem has allowed us to extend and unify previously introduced quantum Schr\"odinger bridges \cite{bergmann1988quantum,pavon2010discrete,georgiou2015positive},   while the problem of bridging density matrices through most likely dynamics, when the basis of the endpoint density matrices is not fixed  (i.e., in the generality of~\cite{georgiou2015positive}), remains. Regardless, the present work constitutes a first step to define and narrow down what is meant by a quantum Schr\"odinger bridge. Undoubtedly, more work is required in this direction, in particular, to understand the connections to other proposed quantum bridges, both from the mathematical \cite{pavon2002quantum,beghi2002steer} and physics \cite{chantasri2013action,chantasri2015stochastic,weber2014mapping} standpoints. Moreover, other so far unrelated quantum problems, whose classical counterparts are a version of the Schr\"odinger bridge problem, have been proposed. 
As was mentioned in the introduction, the classical Schr\"odinger bridge problem can be seen as an energy-minimizing optimal control problem, as well as the entropic regularization of optimal transport. Therefore, related quantum control problems, as well as recently proposed entropic regularizations of non-commutative optimal transport problems \cite{peyre2019quantum,wirth2022dual,feliciangeli2023non}, should be studied in connection to quantum Schr\"odinger bridges to provide an overarching view of what the problem has to offer.

\textit{Acknowledgements.--}
We are grateful to Andrew Jordan for pointing us to the connection between quantum Schr\"odinger bridges and the time-symmetric formulation of quantum measurements. 
OMM would like to thank Luis Correa for his support and insightful discussions. 
OMM was supported by the European Union's Horizon 2020 research and innovation
programme under the Marie Skłodowska-Curie Grant Agreement No. 101151140. RS and TTG were supported in part by the NSF under ECCS-2347357, AFOSR under FA9550-24-1-0278, and ARO under W911NF-22-1-0292.

\bibliography{arXivQSB}

\begin{appendix}

    \subsection*{Appendix: Weak values and their time reversal}
In this Appendix we derive the weak values obtained by weakly measuring an observable $Z=\sum_z z\Pi_z$ in a time-symmetric ensemble with $|x_0^i\rangle$ and $|y_1^j\rangle$ as endpoints.  To do so,
    let us consider a system 
 that evolves according to a Markovian evolution determined by the Kraus operators
    $$
   L_k(1+\delta,0)= K_{l'}(1,\tau)\Omega^{\delta}_{\bar z}K_{l}(\tau,0),
    $$
where $k=\{l,\bar z,l'\}$, $\delta ~>~ 0$, and 
$$
\Omega^{\delta}_{\bar z}={\left(\frac{\delta}{2\pi}\right)^{1/4}}\sum_z e^{-\frac{\delta}{4}(z-\bar z)^2}\Pi_z
.$$
For small $\delta$, this operator describes a weak measurement~\cite{jacobs2006straightforward}. The average value of the outcomes of this generalized measurement, in an ensemble that is pre-selected on state $|x^i_0\rangle$ and post-selected in state $|y^j_1\rangle$, is given by
\begin{align}\nonumber
   & \int_{\mathbb{R}}\bar zP_{\delta}(\bar z|x^i_0,y_1^j)\text{d}\bar{z}=\int_{\mathbb{R}}\bar z\frac{P_{\delta}(\bar z,y_1^j|x^i_0)}{P_{\delta}(y_1^j|x^i_0)}\text{d}\bar{z}
      \\&=\int_{\mathbb{R}}\bar z\frac{\sum_{l,l'}|\langle y_1^j|K_{l'}(1,\tau)\Omega^{\delta}_{\bar z}K_{l}(\tau,0)|x^i_0\rangle|^2}{\int_{\mathbb{R}}\sum_{l,l'}|\langle y_1^j|K_{l'}(1,\tau)\Omega^{\delta}_{\bar z'}K_{l}(\tau,0)|x^i_0\rangle|^2\text{d}\bar{z}'}\text{d}\bar{z}.\label{eq:num-den}
\end{align}The limit of this expression as $\delta\to 0$ will give the weak value
of $Z$, 
$Z_W$.

Specifically, for small $\delta$, 
the measurement operator can be written as
\begin{align}
\Omega^{\delta}_{\bar z}
&=\!\left(\frac{\delta}{2\pi}\right)^{1/4} \!\!e^{-\frac{\delta}{4}\bar z^2}\bigg(I+\delta\frac12 \bar zZ-\delta\frac14 Z^2\!\bigg)\!+O(\delta^2).\label{eq:meas-op}
\end{align} 
With this expression, the denominator in \eqref{eq:num-den} is simply
\begin{align*}
\!\sum_{l,l'}\!|\langle y_1^j|K_{l'}(1,\tau)K_{l}(\tau,0)|x^i_0\rangle|^2\!+O(\delta)\!=\!
    \tr\{\sigma^j_\tau\hat\sigma^i_\tau\}\!+O(\delta),
\end{align*}
 where ${\sigma^j_\tau}$ and $\hat{\sigma}^i_{\tau}$ are given by \eqref{eq:sig} and \eqref{eq:sighat}, respectively. 
 Using \eqref{eq:meas-op} in the numerator of \eqref{eq:num-den}, 
 and  
noting that  
 \begin{align*}
     &\sum_{l,l'}\langle y_1^j|K_{l'}(1,\tau)Z K_{l}(\tau,0)|x^i_0\rangle\langle x^i_0|K_{l}(\tau,0)^\dagger K_{l'}(1,\tau)^\dagger |y_1^j\rangle,\\
     &=\text{tr}(\sigma^j_{\tau}Z\hat{\sigma}^i_{\tau}),
 \end{align*}we obtain,
\begin{align*}
  &  \int_{\mathbb{R}} \bar z\sum_{l,l'}|\langle y_1^j|K_{l'}(1,\tau)\Omega^{\delta}_{\bar z}K_{l}(\tau,0)|x^i_0\rangle|^2\text{d}\bar{z}\\
  &=  \sqrt{\frac{\delta}{2\pi}}\int_{\mathbb{R}} \bar z e^{-\frac{\delta}{2}\bar z^2}\tr\{\sigma^j_\tau\hat\sigma^i_\tau\}\text{d}\bar{z}+O(\delta)
  \\&\quad+\frac{\delta}{2}\sqrt{\frac{\delta}{2\pi}}\int_{\mathbb{R}} \bar z^2e^{-\frac{\delta}{2}\bar z^2}(\tr\{\sigma^j_{\tau} Z\hat{\sigma}_{\tau}^i\} +\tr\{\hat{\sigma}_{\tau}^i Z{\sigma}^j_{\tau}\})\text{d}\bar{z}\\
  &= {\rm Re}(\tr\{\sigma^j_{\tau} Z\hat{\sigma}^i_{\tau}\})+O(\delta),
\end{align*}
where we have used the fact that the mean of the Gaussian distribution is zero, while its variance equals $1/\delta$.
Putting both the numerator and denominator expressions together in \eqref{eq:num-den} and taking the limit as $\delta\rightarrow 0$  yields (the real part of) the weak value 
\begin{align*}
Z^{ij}_W&=\lim_{\delta\rightarrow 0}\int_{\mathbb{R}}\bar zP_{\delta}(\bar z|x^i_0,y_1^j)\text{d}\bar{z} \\& ={\rm Re}\frac{\tr\big\{\sigma^j_\tau Z\hat\sigma^i_\tau\big\}}{\tr\{\sigma^j_\tau\hat\sigma^i_\tau\}},
\end{align*}
as stated in \eqref{eq:weakval}.

To obtain the corresponding time-reversed expression of the weak value, let us consider 
\begin{align*}
     N^{\delta}_{l'}(\tau,1)&:=\rho_{\tau+\delta}^{1/2}K_{l'}(1,\tau)^\dagger\rho_{1+\delta}^{-1/2},
     \\
     \Omega_{\bar z}^{\rm rev,\delta}&:=\rho_{\tau}^{1/2}\Omega^{\delta\dagger}_{\bar z}\rho_{\tau+\delta}^{-1/2},
     \\
N_{l}(0,\tau)&:=\rho_0^{1/2}K_{l}(\tau,0)^\dagger\rho_{\tau}^{-1/2},
\end{align*}
where 
\begin{align*}
 \rho_{\tau}=\sum_lK_l(\tau,0)\rho_0 K_l(\tau,0)^\dagger,~~~~ \rho_{\tau+\delta}=\int_{\mathbb{R}}\Omega^{\delta}_{\bar z}\rho_{\tau}\Omega^{\delta}_{\bar z}\text{d}\bar{z},
\end{align*}are the states of the system before and after the measurement (in the time-forward direction), respectively.

Then, the average value of the outcomes of the generalized measurement in the pre- and post-selected ensemble, determined by $|x_0^i\rangle$ and $|y_1^j\rangle$, is given by
\begin{align*}\nonumber
   & \int_{\mathbb{R}} \bar zP_{\delta}(\bar z|x^i_0,y_1^j)\text{d}\bar{z}=\int_{\mathbb{R}}\bar z\frac{P_{\delta}(\bar z,x^i_0|y_1^j)}{P_{\delta}(x^i_0|y_1^j)}\text{d}\bar{z}
      \\&=\int_{\mathbb{R}} \bar z\frac{\sum_{l,l'}|\langle x^i_0|N_{l}(0,\tau)\Omega^{\delta, \rm rev}_{\bar z}N^{\delta}_{l'}(\tau,1)|y_1^j\rangle|^2}{\int_{\mathbb{R}}\sum_{l,l'}|\langle x^i_0|N_{l}(0,\tau)\Omega^{\delta,\rm rev}_{\bar z'}N^{\delta}_{l'}(\tau,1)|y_1^j\rangle|^2\text{d}\bar{z}'}\text{d}\bar{z}.
\end{align*}
In analogy with equations \eqref{eq:sig}, \eqref{eq:sighat}, we define 
\begin{align*}
    \varsigma^i_\tau&:=\sum_{l}N_{l}(0,\tau)^\dagger|x^i_0\rangle\langle x^i_0|N_{l}(0,\tau),\\
\hat\varsigma^{j}_\tau&:=\sum_{l'}N^{\delta}_{l'}(\tau,1)|y_1^j\rangle\langle y_1^j|N^{\delta}_{l'}(\tau,1)^\dagger.
\end{align*}
Proceeding as in the time-forward direction, we have that 
\begin{align*}
    \int_{\mathbb{R}}\bar zP_{\delta}(\bar z|x^i_0,y_1^j)\text{d}\bar{z} = \cfrac{{\rm Re}(\tr\{\varsigma^i_{\tau} Z^{\text{rev},\delta}\hat{\varsigma}^{j}_{\tau}\rho_{\tau+\delta}^{-1/2}\rho_{\tau}^{1/2}\})}{\tr\{\varsigma^i_\tau\rho_{\tau}^{1/2}\rho_{\tau+\delta}^{-1/2}\hat\varsigma^{j}_\tau\rho_{\tau+\delta}^{-1/2}\rho_{\tau}^{1/2}\}}+O(\delta),
\end{align*}
where 
$Z^{\text{rev},\delta}= \rho_{\tau}^{1/2}Z\rho_{\tau+\delta}^{-1/2}.$ Finally, by noting that  $\rho_{\tau+\delta}\to\rho_\tau$ as $\delta\to 0$  (i.e. in the limit the measurement does not disturb the state), we obtain
\begin{align*}
   Z^{ij,{\rm rev}}_W= &\lim_{\delta \rightarrow 0}\int_{\mathbb{R}}\bar zP_{\delta}(\bar z|x^i_0,y_1^j)\text{d}\bar{z} = 
     {\rm Re}\cfrac{\tr\{\varsigma^i_{\tau} Z^{\text{rev}}\hat{\varsigma}^j_{\tau}\}}{\tr\{\varsigma^i_\tau\hat\varsigma^j_\tau\}} = Z^{ij}_W,
\end{align*}
where 
$$
Z^{{\rm rev}}=\rho_{\tau}^{1/2}Z\rho_{\tau}^{-1/2},
$$
leading to \eqref{eq:weakval-rev} for the case of a more general time-symmetric ensemble.


\end{appendix}

\appendix

\end{document}